\documentstyle[12pt]{article}

\newcommand{\eq}{\begin{equation}}
\newcommand{\en}{\end{equation}}
\newcommand{\eqa}{\begin{eqnarray}}
\newcommand{\ena}{\end{eqnarray}}

\newcommand{\lbl}{\label}

\begin{document}

\hskip 11.5cm \vbox{\hbox{DFTT 15/93}\hbox{April 1993}}
\vskip 0.4cm
\centerline{\bf   TWO DIMENSIONAL QCD}
\centerline{\bf   IS A ONE DIMENSIONAL KAZAKOV-MIGDAL MODEL}
\vskip 1.3cm
\centerline{ M. Caselle, A. D'Adda, L. Magnea  and  S. Panzeri}
\vskip .6cm
\centerline{\sl Istituto Nazionale di Fisica Nucleare, Sezione di Torino}
\centerline{\sl  Dipartimento di Fisica
Teorica dell'Universit\`a di Torino}
\centerline{\sl via P.Giuria 1, I-10125 Torino, Italy}
\vskip 2.5cm

\begin{abstract}
We calculate the partition functions of QCD in two dimensions on a
cylinder and on a torus in the gauge $\partial_{0} A_{0} = 0$ by
integrating explicitly over the non zero modes of the Fourier expansion
in the periodic time variable. The result is a one dimensional
Kazakov-Migdal matrix model with eigenvalues on a circle rather
 than on a line.
We prove that our result coincides with the standard expansion in
representations of the gauge group. This involves a non trivial modular
transformation from an expansion in exponentials of $g^2$ to one in
exponentials of $1/g^2$.
Finally we argue that the states of the $U(N)$ or $SU(N)$ partition
 function can be interpreted as a gas of N free fermions, and the
 grand canonical partition function of such ensemble is given
 explicitly.
\end{abstract}
\vskip 5cm
\hrule
\vskip1.2cm
\noindent

\hbox{\vbox{\hbox{$^{\diamond}${\it email address:}}\hbox{}}
 \vbox{\hbox{ Decnet=(31890::CASELLE,DADDA,MAGNEA,PANZERI)}
\hbox{ internet=CASELLE(DADDA)(MAGNEA)(PANZERI)@TORINO.INFN.IT}}}
\vfill
\eject

\newpage

\section{Introduction}

The last year has seen a revival of attempts to understand
non-perturbative QCD using the $1/N$ expansion .
First, Kazakov and Migdal (KM) \cite{KM} proposed a model with an
 adjoint
multiplet of scalar fields coupled to external gauge fields, which
turned out \cite{Mig} to be exactly solvable in the large $N$ limit.
The model was originally proposed as a way of
finding the master field that would solve four dimensional
QCD in the large $N$ limit. It is not clear yet whether this goal will
be achieved or not, as it is almost certain that some modifications to
the original model are needed for it to admit the QCD fixed point in
the continuum limit, but the proposal has already generated
a considerable literature (see for example \cite{KAZA} and references
therein ).

It has been pointed out \cite{CAPht} that the $d$-dimensional KM
model with a
quadratic potential describes the high temperature limit of pure QCD
in $d+1$ dimensions; however, for $d>1$ such theory does not have a
continuum limit unless higher order terms are added to the potential.
On one hand, this is in agreement with known results in dimensional
reduction of finite temperature QCD, but, on the other hand,
higher terms in the potential are difficult to control
and their arbitrariness limits the predictive power of the model.

Whatever its relationship with four-dimensional QCD,
it is remarkable that the KM model is the only known
example of a matrix model, hence of a string
theory, that is solvable in more than one dimension in the large $N$
limit. The question of what kind of string theory corresponds to the KM
model has also been addressed recently \cite{bou}.
It turns out that the KM model
describes, at least in the large $N$ limit, a string theory with
infinite string tension, where the
string has collapsed into a branched polymer. This might
explain the solvability of the KM model even in dimensions higher than
one.

More recently, in quite a different context, Gross and Taylor
\cite{GT} proved that QCD in two
dimensions (QCD2) can be interpreted as a string theory in the large
 $N$ limit,
by showing that the coefficients of the expansion of the partition
function of QCD2 in power series of $1/N$ can be
interpreted in terms of mappings from a two dimensional surface onto a
two dimensional target space.
A lot of information on the underlying string theory is encoded in the
coefficients of this expansion, but the lack of a prescription for the
string action and for the calculation of the string path integral
limits the efforts to gather non-perturbative information or
extrapolate to higher dimensions.

In this paper, we proceed in a different direction in the understanding
of the string theory underlying QCD2, by proving that QCD2 on
a cylinder and on a torus is described by a matrix model which is
exactly a  one dimensional KM model on a circle with one extra
condition: that each eigenvalue of the scalar field in the KM model be
defined modulo integer multiples of $2 \pi$.
In short, QCD2 on these surfaces is a periodic one-dimensional
KM model, and dimensional reduction (or equivalently the high
temperature expansion) is exact in this case.

We emphasize that we are not taking the large $N$ limit, and our proof
is valid for any $N$. Furthermore, the method we employ leads us to find
expressions for the partition functions that involve exponentials of
the {\it inverse} gauge coupling, unlike the usual character expansions
\cite{Ru,Wit,BlTh}.
The existence of such expressions is a consequence of the well-known
fact that the partition functions of non-abelian gauge theories on
two-dimensional surfaces with boundaries are kernels of the heat
equation on the gauge group manifold, as expected from the path integral
formulation.
Such kernels, at least for $U(N)$ and $SU(N)$, are expected to admit
representations in terms of periodic gaussians in the invariant angles
(sometimes called Weyl angular parameters, they are just $i$ times the
logarithm of the eigenvalues of the unitary matrix in the
fundamental representation). Indeed, such a representation has been
known for a while in the case of the partition function on the disk
\cite{MeOn,dow}.

By a careful gauge-fixing procedure, we are able to find a similar
representation for the kernel associated with the cylinder, and this
leads to the partition function on the torus via identification of the
boundaries and group integration. The invariant angles that parameterize
our representation are precisely the eigenvalues of the KM model, and
are directly related to the Polyakov loops around the cylinder, as
expected from \cite{CAPht}.

After deriving the partition functions, we are able to explicitly
prove the equivalence of our expressions with the corresponding
character expansions, by performing a modular transformation that
generalizes to non-abelian gauge theories the modular inversion of
Jacobi's $\theta_3$, which in fact is the partition function of QED on a
torus.

As a result of our analysis we also find a generating function for the
partition functions of $U(N)$ [$SU(N)$] . As expected from the
identification of QCD2 with a one dimensional matrix model, this grand
canonical partition function is given in terms of infinite products
of free fermion partition functions.

The paper is organized as follows: in section $2$, after a few general
remarks, we calculate the partition function on the cylinder with an
appropriate choice of gauge fixing, and we get a representation in terms
of angles that coincides with the announced KM model. In section $3$ we
proceed to reconstruct the character expansion on the cylinder by a
modular transformation. In section $4$ we calculate the partition
function on the torus and we find the generating function that reproduces
the partition functions of $SU(N)$ ($U(N)$) for any $N$. In section
$5$ we present our conclusions, and finally in the appendix we present
 an alternative derivation of the kernel
on the cylinder, following closely the techniques applied to
hermitian matrix models in \cite{IZ}.

\section{The Partition Function on the Cylinder}

It is well known by now that two dimensional QCD defined on a manifold
$\cal M $ of genus $G$ and with a metric $g_{\mu \nu} $ is exactly
solvable. The partition function is given by
\eqa
{\cal Z}_{\cal M}(N, {\cal A}) & = & \int {\cal D} A_{\mu}
 e ^{- \frac{1}{4 \tilde{g}^{2}}
\int_{\cal M} d^{2} x \sqrt{g} {\rm Tr} F_{\mu \nu}
F^{\mu \nu}} \nonumber \\
 & = &
\sum_{R} d_{R}^{2-2G} e^{- \frac{1}{2} {\cal A} \tilde{g}^{2} C_{2}(R)} ,
\label{eq1}
\ena
where the sum is over all equivalence classes of irreducible
representations $R$,
$d_{R}$ is their dimension and $ C_{2}(R) $ is
the quadratic Casimir in the representation $R$.
Similarly, the heat kernel defined by a surface of genus $G$ and
$n$ boundaries is given by \cite{Wit,BlTh}
\eq
{\cal K}_{G,n} (g_1, \ldots , g_n; N, {\cal A}) = \sum_R
d_R^{2 - 2G - n} \chi_R(g_1) \cdots \chi_R(g_n)
e^{-\frac{1}{2} {\cal A} \tilde{g}^2 C_2(R)} ,
\label{heatker}
\en
where $g_i$ are the Wilson loops evaluated along the boundaries, and
$\chi_R$ denotes the Weyl character of the representation $R$.
For dimensional reasons, and because of the invariance of the action
under area preserving diffeomorphisms, eqs. (\ref{eq1}) and
 (\ref{heatker})
depend only on the variable  $\tilde{g}^2 {\cal A}$.
We will henceforth denote this variable by $t$.

The heat kernels in eq. (\ref{heatker}) are class functions,
 and therefore
they must admit a representation in terms of the eigenvalues of the
group elements $g_i$ in the fundamental representation, which are
pure phases for $SU(N)$ and $U(N)$. An explicit expression in terms of
 these
phases was found by Menotti and Onofri \cite{MeOn} for the simplest
case, the disk. It is of the form
\eqa
{\cal K}_{0,1}(\phi_i, t) & = &{\cal N}(t) \sum_{\{l_i\} = - \infty}^{+
\infty}
\prod_{i<j=1}^N \frac{\phi_i - \phi_j + 2 \pi (l_i - l_j)}{2 \sin
\frac{1}{2}\left[\phi_i - \phi_j + 2 \pi (l_i - l_j)\right]} \nonumber
\times \\
& \times &
\exp \left[-\frac{1}{t} \sum_{i = 1}^N \left(\phi_i + 2 \pi l_i\right)^2
\right] .
\lbl{dowmeon}
\ena

In order to formulate QCD2 as a KM model, we will construct
a similar representation for the kernel defined by the cylinder
(${\cal K}_{0,2}$), which we will denote simply by ${\cal K}_2$.
It will be convenient to work in first order
formalism, where the action is given by
\eq
S(N,t) = \frac{2}{t} \int_0^{2\pi}
dx d\tau~ {\rm Tr} [F^2 - i F f(A)] ,
\label{eq2}
\en
where
\eq
f(A) = \partial_0 A_1 - \partial_1 A_0 -i[A_0,A_1] ,
\label{eq3}
\en
while $F$ and $A$ are considered as independent fields.
By using the invariance under area preserving diffeomorphisms of the
original action, we have restricted the metric in eq. (\ref{eq3})
to be flat
and rescaled space-time coordinates $(x, \tau)$ so that they both
 range in the interval $(0, 2 \pi)$. The variable $t$ appears then
explicitly at the r.h.s.

In the case of a cylinder we have to fix the value of the Polyakov
loop at the two boundaries (say at $x=0$ and $ x=2\pi $)
by introducing suitable delta functions :
\eqa
{\cal K}_2 ( g_1,g_2, t) & = &
\int {\cal D} A_\mu {\cal D} F
\exp \left(-\frac{2}{t} {\rm Tr} \int_0^{2 \pi}
dx d\tau [ F^2 - i F f(A) ] \right) \times
\nonumber \\
& \times &
\hat{\delta} \left(W(0),g_1\right) \hat{\delta} \left(W(2 \pi),g_2\right)
\psi(g_1) \psi(g_2) ,
\label{eq4}
\ena
where
\eq
W(x) = {\cal P} e^{i \int_0^{2\pi} d\tau A_0(x,\tau)} ,
\label{eq5}
\en
and $\hat{\delta}(g,h)$ denotes the conjugation invariant delta function
on the group manifold, defined by
\eq
\hat{\delta} (g,h) = \int dU \delta (U g U^\dagger h) .
\lbl{delta}
\en
The factors $ \psi(g_1) $ and $ \psi(g_2)$ are just normalization
factors; they depend only on the eigenvalues of $g_1$ and $g_2$
and they will be chosen so that the
sewing of two cylinders corresponds to just a group integration:
\eq
\int {\cal K}_2 (g_1,g,) {\cal K}_2 (g^{\dagger},g_2) dg
= {\cal K}_2 (g_1,g_2) .
\label{eq6}
\en
All fields in eq. (\ref{eq4}) are periodic with period $2 \pi$ in $\tau$
and can be expanded in Fourier series as
\eq
\left[
\begin{array}{c}
A_{0}(x,\tau) \\
A_{1}(x,\tau) \\
F(x,\tau)
\end{array}
\right]
= \sum_{n=-\infty}^{+\infty} \left[
\begin{array}{c}
B_{n}(x) \\  A_{n}(x) \\ F_{n}(x) \end{array}
\right]
e^{i\,n\,\tau} .
\label{eq7}
\en
All the Fourier modes with $n \neq 0 $ can be integrated away in
eq. (\ref{eq4}) with a suitable gauge choice, namely
\eq
\partial_{0} A_{0}(x,\tau) = 0 ~~ \leftrightarrow
{}~~ B_{n}(x) = 0, \quad n \neq 0 .
\label{eq8}
\en
In this gauge the kernel becomes
\eqa
{\cal K}_2 (g_1, g_2, t) & = & \int {\cal D} B_{0} \prod_{n}
{\cal D} A_{n} {\cal D} F_{n}
e^{-\frac{1}{t} {\rm Tr} \int_{0}^{2\pi}
dx {\cal L}(B_{0},A_{n},F_{n}) } \Delta_{FP} \times \nonumber \\
& \times &
\hat{\delta} \left(W(0),g_1\right) \hat{\delta} \left(W(2 \pi),g_2\right)
\psi(g_1) \psi(g_2) ,
 \label{eq9}
\ena
where
\eq
{\cal L} = \sum_{n=-\infty}^{+\infty} \{
F_{n}F_{-n} - n F_{n} A_{-n} + i\delta_{n,0}F_{0}\partial_{1}B_{0}
- F_{n} [B_{0},A_{-n}] \} ,
\label{eq10}
\en
while $ \Delta_{FP} $ is the Faddeev-Popov determinant
\eq
\Delta_{FP} = \prod_{n \neq 0} det
\frac{\delta [ n \varepsilon_{n} - [B_{0},\varepsilon_{n}]]_{rs}
}{\delta\varepsilon_{n,tl}}
\label{eq11}
\en
and $ W(x) $ is now given by
\eq
W(x) = e^{2i\pi B_{0}(x)} .
\label{eq12}
\en
At this point the functional integration over $ A_{n}~ (n \neq 0) $
leads to a product of delta functions :
\eq
\prod_{n \neq 0} \delta \left( nF_{n} - [B_{0},F_{n}] \right) =
\frac{1}{\Delta_{FP}} \prod_{n \neq 0} \delta(F_{n}) .
\label{eq13}
\en
By inserting this equation in eq. (\ref{eq9}) we obtain
\eqa
{\cal K}_2(g_{1},g_{2}) & = &
\int {\cal D} B {\cal D} A {\cal D} F
e^{-\frac{4\pi}{t} {\rm Tr} \int_{0}^{2\pi}
dx \left( F^{2} + iF\partial B - F[B,A] \right) } \times \nonumber \\
& \times &
\hat{\delta} \left(W(0),g_1\right) \hat{\delta} \left(W(2 \pi),g_2\right)
\psi(g_1) \psi(g_2)
\nonumber \\
 & = & \int {\cal D}B {\cal D}A e^{-\frac{\pi}{t}
{\rm Tr} \int_{0}^{2\pi} dx \left[ \partial B - i[A,B] \right]^{2}}
\times \nonumber \\
& \times &
\hat{\delta} \left(W(0),g_1\right) \hat{\delta} \left(W(2 \pi),g_2\right)
\psi(g_1) \psi(g_2) .
\label{eq14}
\ena
where we have replaced $B_0,A_0 $ and $F_0$ with $B,A$ and $F$.
This result can be interpreted as follows: the action is a KM model in
one continuum dimension with boundary conditions depending on
$ e^{2i\pi B} $ rather than $ B $;  this implies that if, for instance,
$ B(0) $ satisfies the boundary conditions imposed by the delta
functions at $ x = 0 $, then the latter are satisfied by any other
matrix whose set of eigenvalues coincides
with the one of $B(0)$ {\it modulo
integer numbers}. It is indeed possible to restrict all the
eigenvalues of $ A_{0}(x,\tau) = B(x) $ to be in the interval $[0,1)$
by a further gauge choice. In fact if
$ A_{0}(x,\tau)  = U(x){\rm diag} \left( \lambda_{i}(x) \right)
 U^{\dagger}(x) $ then the gauge transformation generated by the
 unitary matrix
$ h(x,\tau) = U(x) {\rm diag} \left( e^{i n_{i} \tau } \right)
 U^{\dagger}(x) $
is periodic in $\tau$ with period $2\pi$, preserves the gauge choice
(\ref{eq8}), and  just amounts to a shift of all the eigenvalues
$\lambda_{i}(x)$ by the integer numbers $ n_{i} $.
It is clear though that to avoid discontinuities in the gauge
transformation the integers $ n_{i} $ in $ h(x,t) $ have to be $x$
independent , so the eigenvalues $ \lambda_{i}(x) $ can be restricted to
the interval $ [0,1) $ only for one particular value of $x$, say $ x = 0
$.
By means of techniques which are by now standard in matrix models, the
matrix $ B $ can be diagonalized and the action written in terms of its
eigenvalues $ \lambda_{i}(x) $, as
\eqa
{\cal K}_2(g_{1},g_{2},t) & = & \int {\cal D} \lambda_{i}(x)
\Delta \left( \lambda(0) \right)
\Delta \left( \lambda(2\pi) \right)
\exp \left\{ - \frac{\pi}{t}
\int_{0}^{2\pi} dx \partial \lambda_{i}(x) \partial \lambda^{i}(x)
\right\} \times
\nonumber \\
& \times &
\int dU_{1} dU_{2} \delta \left[ U_{1} e^{2i\pi \lambda(0)}
U_{1}^{\dagger} e^{-i\theta} \right]
\delta \left[ U_{2} e^{2i\pi \lambda (2\pi)} U_{2}^{\dagger}
e^{-i\phi} \right] \psi(\theta) \psi(\phi) ,
\label{eq15}
\ena
where we inserted the explicit expression for the $\delta$ functions,
we denoted by $\lambda(0)$, $\lambda(2\pi)$, $\theta$, $\phi$
the diagonal matrices whose eigenvalues are respectively
$ \lambda_{i}(0)$,
$\lambda_{i}(2\pi)$, $\theta_{i}$, and $\phi_{i}$,
and we chose $ e^{- i\theta} $
[resp $ e^{- i\phi} $] to denote
the diagonal form of $g_{1}$ [resp of $g_{2}$].
$\Delta (\lambda)$ is the Vandermonde determinant of a hermitian matrix
with eigenvalues $\lambda_i$.

The integrals over the unitary matrices $ U_{1} $ and $ U_{2} $ can be
calculated from the identity
\eqa
\exp \left[\beta {\rm Tr}(V + V^\dagger - 2) \right] & = &
\sum_R d_R \beta^{- N^2/2} \chi_R(V) \left( 1 + O(1/\beta) \right)
\nonumber \\
& = & \beta^{- N^2/2} \delta(V) ,
\lbl{saddle}
\ena
where $V$ is a unitary matrix.

Inserting this expression of $\delta(V)$ in eq. (\ref{delta}),
and using the saddle point method to calculate
the integral for large $\beta$\footnote{These generalizations of the
Harish-Chandra integral are discussed in section $3$ of
ref. \cite{KMSW}}, we find
\eq
\int dU \delta \left(U e^{i \sigma} U^\dagger e^{i \xi} \right) =
 \sum_P \sum_{\{n_j\} = - \infty}^{+ \infty}
 \frac{(-1)^{P + (N-1) \sum_k n_k}}{J(\sigma)
J(\xi)}
\prod_{j = 1}^N \delta(\sigma_j + \xi_{P(j)} + 2 \pi n_j) ,
\lbl{intdelta}
\en
where $P$ denotes a permutation of indices,
\eq
J(\sigma) = \prod_{i<j} 2 \sin \frac{\sigma_i - \sigma_j}{2}
\lbl{vandermonde}
\en
is the Vandermonde determinant for a unitary matrix, and we made use of
the fact that $J(\sigma + 2 \pi n) = (-1)^{(N-1) \sum_j n_j} J(\sigma)$.

The quadratic functional integral over the eigenvalues can be performed
by a $\zeta$-function regularization of the divergences, leading to
\eqa
& &\int  {\cal D} \lambda_i(x) \exp \left[ - \frac{\pi}{t}
\int_0^{2 \pi} dx \sum_{i = 1}^N \left( \partial \lambda_i(x) \right)^2
\right] \nonumber \\ & &
= \left( \frac{1}{4 \pi^2 t} \right)^{N/2}
\exp \left[ - \frac{1}{2t} \sum_{i = 1}^N \left(\lambda_i(0) -
\lambda_i(2 \pi) \right)^2 \right]  .
\lbl{kazmig}
\ena

The result at the r.h.s. of eq. (\ref{kazmig}) is the same that we would
have obtained if we had started from a KM model on a one dimensional
lattice with an arbitrary number of sites in the interval $[0, 2 \pi)$.
The result is independent
of the number of sites due to the remarkable scaling properties
of the KM model, already noticed in \cite{CAP}.
By substituting eq. (\ref{kazmig}) and eq. (\ref{intdelta}) into
eq. (\ref{eq15}) one finally obtains
\eqa
{\cal K}_2(g_1,g_2,t) & = & \sum_P
       \frac{t^{-N/2}}{J(\theta) J(\phi)}
       \sum_{\{l_{i}\}}  (-1)^{P + (N-1) \sum_j l_j} \nonumber \\
 & &   \exp \left[ - \frac{1}{2t} \sum_{i = 1}^N
       \left( \phi_i - \theta_{P(i)} + 2 \pi l_i \right)^2     \right] .
\lbl{kcyl}
\ena
Here the normalization factor has also been determined. It is
given by $\psi(g) = J(g)/\Delta(g)$, where $\Delta(g) =
\prod_{i<j} (\theta_i - \theta_j)$ if the eigenvalues of $g$ are denoted
by $e^{i \theta_i}$.

For the gauge group $U(N)$ the eigenvalues $\phi_i$, $\theta_i$ and the
integers $l_i$ are unconstrained, whereas for $SU(N)$ we can choose
$\sum_i \phi_i = \sum_i \theta_i = 0$, and this in turn constrains the
integers $l_i$ to obey $\sum_i l_i = 0$. In fact if we choose the gauge
$\lambda_i(0) = \theta_i$, as previously discussed, then the fact that
$\lambda_i(2 \pi) = \phi_i + 2 \pi l_i$, together with the continuity
of the function
$\sum_i \lambda_i(x)$, implies $\sum_i l_i = 0$.
Thus for $SU(N)$ the sign factor
in each term of eq. (\ref{kcyl}) is just given by $(-1)^P$.
 Note also that for $SU(N)$ the
factor $t^{-N/2}$ must be replaced by $t^{(1-N)/2}$.

\section{Modular Inversion of the Kernel on the Cylinder}

The expression for the kernel on a cylinder obtained in the
previous section, as announced, involves exponentials
in $1/t$, unlike eq. (\ref{heatker}).
As a preliminary check on eq. (\ref{kcyl}),
one can easily verify that by taking the limits $\theta
 \rightarrow 0 $
or $\phi \rightarrow 0 $ one obtains the kernel on the
disk, eq. (\ref{dowmeon}).

To verify that the two expressions are related by a modular
transformation, let us consider now eq. (\ref{kcyl}) for
the gauge group $SU(N)$ :
\eqa
{\cal K}_2(g_1,g_2,t) & = & \frac{1}{2 \pi N!} \sum_{P,P^\prime}
       \frac{t^{(1-N)/2}}{J(\theta)
       J(\phi)}\sum_{\{l_{i}\}} \int_0^{2 \pi}
       d\beta (-1)^{P + P^\prime} \nonumber \\
 & &   \exp \left( - \frac{1}{2t} \sum_{i = 1}^N
       \left[ \left( \phi_{P^\prime(i)} - \theta_{P(i)} +
       2 \pi l_i \right)^2 - 2i \beta t  l_i \right] \right) .
\label{newK2}
\ena
The lagrange multiplier $\beta$ has been introduced to impose the
 condition $\sum l_i = 0$, and a redundant double sum over
 permutations has replaced the simple sum of eq.~(\ref{kcyl}).

By completing the square, eq. (\ref{newK2}) can be rewritten as
\eqa
{\cal K}_2(g_1,g_2,t) & = & \frac{1}{2 \pi N!} \sum_{P,P^\prime}
       \frac{t^{(1-N)/2}}{J(\theta) J(\phi)}
       \sum_{\{l_i\}} \int_0^{2 \pi} d\beta (-1)^{P + P^\prime}
       \nonumber \\
 & &   \exp \left( - \frac{2 \pi^2}{t} \sum_{i = 1}^N
       \left( l_i + \frac{\phi_{P^\prime(i)} - \theta_{P(i)}}{2 \pi}
       - i \frac{\beta}{4 \pi^2} t \right)^2 \right) \times \\
 & \times & \exp \left( - \frac{N t \beta^2}{8 \pi^2} \right) ,
 \nonumber
\label{complsquare}
\ena
so that one can use the well known modular transformation of the function
$\theta_3$,
\eq
\sum_{l = - \infty}^{+ \infty} \exp \left(-\frac{(\theta + l)^2}{4 t}
 \right)
      = \sum_{n = - \infty}^{+ \infty} \exp \left( - 4 \pi^2  n^2 t
 \right)
      \exp (2 \pi i n \theta) (4 \pi t)^{1/2} ,
\label{modtransf}
\en
to obtain
\eqa
 & &  {\cal K} (g_1,g_2,t) = \frac{1}{2 \pi N!} \sum_{P,P^\prime}
       \frac{t^{1/2}}{J(\theta) J(\phi)}
       \sum_{\{n_i\}} \int_0^{2 \pi} d\beta  (-1)^{P + P^\prime}
       \times         \nonumber \\
 & & \times  \exp \left[ - \frac{t}{2} \sum_{i = 1}^N n_i^2
       + 2 \pi i \sum_{i = 1}^N n_i
       \left( \frac{\phi_{P^\prime(i)} - \theta_{P(i)}}{2 \pi}
       - i \frac{\beta}{4 \pi^2} t \right)
       - \frac{N t \beta^2}{8 \pi^2} \right]  \nonumber \\
 & & = \frac{t^{1/2}}{2 \pi N!} \sum_{\{n_i\}}
       \exp \left(- \frac{t}{2} \left[
       \sum_i n_i^2 - \frac{1}{N} \left(\sum_i n_i \right)^2
       \right] \right) \times \nonumber \\
 & & \times \frac{ \det \{ e^{i n_i \phi_j} \} }{J(\phi)}
       \frac{ \det \{ e^{- i n_i \theta_j} \} }{J(\theta)} \times \\
 & & \times   \int_0^{2 \pi} d\beta
       \exp \left[ - \frac{N t}{8 \pi^2}
       \left(\beta - \frac{2 \pi}{N} \sum_i n_i \right)^2 \right] .
 \nonumber
\label{bigarray}
\ena

The sum over $\{n_i\}$ can now be restricted to the region $n_1 > n_2 >
\cdots > n_N$, at the expense of the factor $1/N!$.
 One can further notice that both the first exponential and the
 determinants are invariant if all the
$n_i$'s are shifted by the same constant. One is then lead to define
$r_i = n_i - n_N - N$, in terms of which the kernel becomes
\eqa
{\cal K}_2(g_1, g_2, t) & = & \frac{\sqrt{t}}{2\pi}
\sum_{r_1 > r_2 > \ldots > r_N = -N}
\exp \left\{ -  \frac{t}{2} \left[ \sum_{i} r_{i}^2
- \frac{1}{N} \left( \sum r_{i} \right)^2 \right] \right\} \times
\nonumber \\
   & \times & \frac{ \det \{ e^{i r_i \phi_j} \} }{J(\phi)}
       \frac{ \det \{ e^{- i r_i \theta_j} \} }{J(\theta)} \times \\
   & \times &    \sum_{n_N = - \infty}^{+ \infty}
       \int_{\beta_0}^{\beta_1} d\beta \exp
       \left[ - \frac{N t}{8 \pi^2}
       \left(\beta - \frac{2 \pi}{N} \sum_i r_i \right)^2
       \right] , \nonumber
\lbl{pippo}
\ena
where $\beta_0 = - 2 \pi n_N - 2 \pi N = \beta_1 - 2 \pi$.
The sum over $n_N$ simply reconstructs the gaussian integral from
$- \infty$ to $+ \infty$, that can be trivially calculated. It is
easy to recognize now that in eq. (28) the exponential is
related to the quadratic Casimir of the representation $R$ whose Young
tableaux has rows of length $\hat{r}_i = r_i + i - N$. In fact
\eq
\sum_i r_i^2 - \frac{1}{N} \left( \sum_i r_i \right)^2 = C_2 (R) +
\frac{1}{12} N (N^2 -1) .
\lbl{casimir}
\en
The last term can be interpreted as a zero point energy, namely the
energy of the lowest representation. It is proportional to the scalar
curvature of the group manifold, and to the modulus squared of the
vector obtained by summing the positive roots of $SU(N)$.
On the other hand we have also
\eq
\chi_R(\phi) = \frac{\det \{e^{i r_i \phi_j}\}}{J(\phi)}
(-i)^{\frac{N(N-1)}{2}}  .
\lbl{char}
\en
By substituting  (\ref{casimir}) and  (\ref{char}) into eq.
(28) we obtain the modular inversion for the $SU(N)$ kernel on
 the cylinder, which reads
\eqa
& & \exp \left( - \frac{t}{24} N (N^2 -1) \right)
\sum_R \exp \left( - \frac{t}{2} C_2 (R) \right)  \chi_R (-\theta)
\chi_R(\phi) \nonumber \\
& = &  \left( \frac{N}{4 \pi} \right)^{1/2}
\sum_P \frac{t^{(1-N)/2}}{J(\theta)
 J(\phi)} \sum_{\{l_{i}\}}
  \exp \left[ - \frac{1}{2t} \sum_{i = 1}^N
       \left( \phi_i - \theta_{P(i)} + 2 \pi l_i \right)^2     \right] ,
\lbl{modinv}
\ena
where the constraints $\sum_i l_i  = \sum_i \phi_i = \sum_i
\theta_i = 0$ are understood.

\section{The Partition Function On the Torus}

The kernel on a cylinder obtained in section 2
section allows one to calculate the partition function
of QCD2 on a torus by simply sewing together the two ends of
the cylinder, according to
\eq
{\cal Z}_{G=1}(N,t) = \int dg {\cal K}_{2}(g,g,t) =
\int_{0}^{2\pi} \prod_i d\phi_i J^{2} (\phi) {\cal K}_{2}(\phi,\phi,t) .
\en
This was done in ref. \cite{BlTh} by using eq. (\ref{heatker}) with
 $G=0$, $n=2$ and the orthogonality
properties of the characters. By repeating the same calculation with our
expression (\ref{kcyl}) for the kernel we are able to write
${\cal Z}_{G=1}(N,t)$ in terms of $\theta$  functions, whose
behaviour under modular transformations is well known.
The result is particularly simple in the case of the group $ U(N) $ where
in the expression of the kernel, eq. (\ref{kcyl}), the integers $ l_{i} $
and the angles $ \theta_{i} $ and $ \phi_{i} $ are unconstrained.

In this case the partition function is given by \footnote{This partition
 function has a different zero point energy compared to
 the  one defined for instance in ref. \cite{BlTh} corresponding to
the overall factor $ \exp [ \frac{t}{2} \frac{N(N^{2}-1)}{12}] $}
\eqa
{\cal Z}_{G=1}(N,t) & = & \left(\frac{t}{4\pi} \right)^{-N/2}
\int_{0}^{2\pi} \prod_{i = 1}^N d\phi_i \sum_{P}
(-1)^{P} \times \nonumber \\
& \times & \sum_{\{l_{i}\}} (-1)^{(N-1) \sum_j l_j}
 \exp \left(- \frac{1}{2t} \sum_{i=1}^{N}
 ( \phi_{i} - \phi_{P(i)} + 2 \pi l_{i} )^2 \right) .
\label{torus1}
\ena

By using the modular transformation eq. (\ref{modtransf})
for the theta function
the exponentials can be made linear in the angles $ \phi_{i} $ :
\eqa
{\cal Z}_{G=1}(N,t) & = & \int_{0}^{2\pi} \prod_{i = 1}^N d\phi_i
\sum_{P} (-1)^{P} \times \nonumber \\
& \times & \sum_{\{n_{i}\}}
\exp \left( - \frac{t}{2} \sum_{i=1}^{N} \left(n_i - \delta_N\right)^{2}
+ i \sum_{i} \left(n_i - \delta_N\right) (\phi_{i} - \phi_{P(i)} )
\right) ,
\lbl{linear}
\ena
where $\delta_N$ is $0$ for odd $N$ and $1/2$ for even $N$.
The r.h.s. of eq. (\ref{linear}) can be computed by the same
method employed in
\cite{CAP} for the one-dimensional KM model, that is by decomposing each
permutation into its cycles. Then
\eq
{\cal Z}_{G = 1}(N,t) = \sum_{h_1, \ldots , h_N} \delta \left(
\sum_{r = 1}^N r h_r - N \right) (-1)^{\sum_j (j-1) h_j}
\prod_{r = 1}^N \left(\frac{F_r}{r}\right)^{h_r} \frac{1}{h_r!} ,
\lbl{cycles}
\en
where $ h_r $ is the multiplicity of a cycle of length $r$ in a given
permutation and the sum over all permutations is reproduced by summing
over the $ h_r $ 's with the correct combinatorial factors. $ F_r $ is
the contribution from a cycle of length $r$ and  it is given by
\eqa
F_r & = & \sum_{\{n_i\}} \int_{0}^{2\pi} d\phi_1 \ldots d\phi_r
\exp \left[ -\frac{t}{2} \sum_{i=1}^r \left(n_i - \delta_N\right)^{2} +
i\sum_{i=1}^{r} \left(n_i - \delta_N\right)
\left( \phi_i - \phi_{i+1} \right) \right]
\nonumber \\
& = & (2\pi)^{r} \sum_{n= -\infty}^{+ \infty}
e^{-\frac{t~r}{2} \left(n - \delta_N\right)^2}
= (2\pi)^r \theta_{\sigma(N)} \left(0,\tau = i\frac{t r}{2\pi} \right) ,
\label{F_r}
\ena
where $\theta_{\sigma(N)}$ denotes Jacobi's $\theta_2$ for $N$ even and
$\theta_3$ for $N$ odd.

This result can be expressed in a rather elegant and interesting form if
one consider the grand-canonical partition function
\eq
{\cal Z}_{G=1}(q) = \sum_N {\cal Z}_{G=1}(N,t) q^N
\label{grpartU}
\en
The even and odd parts of ${\cal Z}_{G=1}(q)$ have to be computed
separately. From eq. (\ref{grpartU}) we have
\eqa
{\cal Z}^{{\rm even}}_{G=1}(q) & = &
\frac{1}{2} \exp \left\{ \sum_{r=1}^{\infty} \frac{(-1)^r}{r}
\theta_2 \left( 0, \tau = i\frac{t~ r}{2\pi}
\right) ( 2 \pi q)^r \right\} + \left( q \leftrightarrow -q \right) ,
\nonumber \\
{\cal Z}^{{\rm odd}}_{G=1}(q) & = &
\frac{1}{2} \exp \left\{ \sum_{r=1}^{\infty} \frac{(-1)^r}{r}
\theta_3 \left( 0, \tau = i\frac{t~ r}{2\pi}
\right) ( 2 \pi q)^r \right\} - \left(q \leftrightarrow -q\right) .
\label{Noddeven}
\ena

The sum over $r$ in the exponents can be performed if one replaces the
theta functions with their expressions as infinite sums, finally leading
to the result
\eqa
{\cal Z}^{{\rm even}}_{G=1}(q) & = & \frac{1}{2} \prod_{n=0}^{\infty}
\left( 1 + 2\pi q e^{-\frac{t}{2} \left( n + \frac{1}{2} \right)^2}
\right)^2 + \left( q \leftrightarrow -q \right) , \nonumber \\
{\cal Z}^{{\rm odd}}_{G=1}(q) & = &
\frac{1}{2} (1 + 2 \pi q) \prod_{n=1}^{\infty}
\left( 1 + 2\pi q e^{-\frac{t}{2} n^2} \right)^2
- \left( q \leftrightarrow -q \right) .
\label{gasfermions}
\ena

This partition function has some similarity with the one of the
one-dimensional KM model \cite{CAP}, in
the sense that they both describe a
gas of free fermions. However, in this case there are some
peculiarities. The integers $n$ can be interpreted as the discretized
momenta associated with the winding of the eigenvalues, therefore the
energy levels are proportional to $n^2$, rather than $n$.
For $ n\neq 0 $ two degenerate levels are present,
corresponding to winding in opposite directions.
On the other hand, the integers $l_i$ in eq. (\ref{torus1}) are the
winding numbers, namely  the discretized coordinates of such modes, and
the modular transformation allows us to go from the coordinate to the
momentum representation.
Further, the energy levels depend on the parity of the
total number of fermions, which is connected with the fact that a
fermionic wave function picks up a different sign according to the
number of other fermions encountered in the winding. This is
the meaning of the phase factor $(-1)^{(N-1) \sum_i l_i}$ in eq.
(\ref{kcyl}).

We also emphasize that ${\cal Z}_{G=1}(q)$ has been defined
keeping $t=\tilde{g}^{2} {\cal A}$ independent of $N$, so eq. (
\ref{gasfermions}) cannot be used as such to calculate the large $N$
limit of ${\cal Z}_{G=1}(N,t)$.

At least for $N$ odd, eq. (\ref{Noddeven}) has a simple connection with
QED2, as one can see by remembering that
 $\theta_3 (0,i~t/2\pi) $ is the partition function of QED2 on a
surface of area ${\cal A}$ \cite{BlTh}.
Consider now QCD2 on a torus and go into the gauge where $ F(x,t) $ is
diagonal. An easy calculation shows that the Vandermonde determinants
resulting from diagonalizing $ F(x,\tau) $ are exactly cancelled by the
functional integral over the off-diagonals elements of $ A_{\mu}(x,\tau)
$.
 This leaves one apparently with $N$ copies of QED. However,
 as one goes
with continuity from $ x = 0 $ to $ x = 2\pi$ the eigenvalues of $F$
will be permuted in an arbitrary way, and the sum over all permutations
is needed to recover the whole partition function. One should notice
finally that, in the decomposition of each permutation into irreducible
cycles, each cycle of length $r$ corresponds to a QED2 where we have to
wind $r$ times in the $x$ direction before we go back to the original
point, so in fact a QED2 on a torus of area $ r {\cal A} $.

Consider now the case of $SU(N)$. The partition function for the group $
SU(N)$ is the same as eq. (\ref{torus1}), but with $N$ replaced by $N-1$
 and the constraint $ \sum \phi_i = \sum \l_i = 0 $. Thus
\eqa
{\cal Z}_{G=1}^{SU(N)}(N, t) & = & \left( \frac{t}{4\pi}
\right)^{\frac{1-N}{2}}  \int_{0}^{2\pi} d\phi_1 \ldots d\phi_N
\delta \left[ \sum_i \phi_i \right] \sum_P (-1)^P \int_{0}^{2\pi}
\frac{d\beta}{2\pi} \sum_{\{l_i\}}
\nonumber \\
& & \times
\exp \left\{ - \frac{1}{2t} \sum_{i=1}^{N}
\left( \phi_i - \phi_{P_i} + 2 \pi l_i \right)^2 + i\beta \sum_{i=1}^{N}
l_i \right\} .
\ena
The sum over the $ l_i $ can be done by reconstructing the square and
using eq. (\ref{modtransf}), in analogy to what was done in in the case
 of kernel
of the cylinder. At this point the integral over the $ \phi_i $ is
trivial and gives
\eqa
{\cal Z}_{G=1}^{SU(N)}(N, t) & = & \left( \frac{t}{4\pi}
\right)^{1/2} \sum_{\{n_i\}} \sum_P (-1)^P \prod_{i=1}^{N}
\delta_{n_i n_{P_{i}}}
\exp \left\{ - \frac{t}{2} \left[ \sum_{i=1}^{N}
n_{i}^{2} - \frac{(\sum n_i)^2}{N} \right] \right\}
\nonumber \\ & \times &
\int_{0}^{2\pi} d\beta \exp \left\{ -\frac{Nt}
{8 \pi ^2} \left(\beta - \frac{2\pi}{N} \sum n_i \right) \right\}.
\label{SUNpart}
\ena
It is not difficult now to show that this partition function coincides
(apart from the zero point energy) with the one of ref.
 \cite{Ru,Wit,BlTh},
\eq
{\cal Z}_{G=1}(N, t) = \sum_R e^{- \frac{t}{2} C_2 (R)}.
\label{partition}
\en
In fact the determinant of $ \delta $ functions in eq. (\ref{SUNpart})
 just
forbids any couple of $ n_i$'s to take the same value and the integral
over $ \beta$ becomes trivial after the shift of the $n_i$'s leading
 to  eqs. (27) and (28).

Starting from eq. (\ref{SUNpart})
 it is possible, following the steps leading to eqs. (\ref{grpartU}) and
 (\ref{gasfermions}), to write in compact form the
grand canonical partition function for $SU(N)$, as
\eqa
{\cal Z}_{G=1}^{SU(N)} (q)&  = & \sum_N {\cal Z}_{G=1}^{SU(N)}(N, t) q^N
\nonumber \\
& = & \left( \frac{t}{4\pi} \right)^{1/2} \int_{0}^{2
\pi}
d\beta \prod_{n=-\infty}^{+\infty} \left( 1 + q
e^{- \frac{t}{2} \left( n - \frac{\beta}{2\pi}
\right)^2 } \right) .
\label{gasfermionsSUN}
\ena
The interpretation of this formula is the following: the constraint
$\sum_i l_i = 0$ means that for $SU(N)$ the wave function of the center
of mass of the eigenvalues is completely localized, and therefore the
corresponding momentum is undetermined. This is the reason of the
integration over $\beta$ in eq. (\ref{gasfermionsSUN}), and of the
invariance of the Casimir in eq. (\ref{casimir}) under common shifts of
the integers labelling the representations.

\section{Conclusions}

We have shown in this paper, by means of dimensional reduction
techniques, that QCD2 on a cylinder or a torus is exactly a one
dimensional matrix model of the type proposed by Kazakov and Migdal,
with the substantial new feature that the eigenvalues of the matter
fields live on a circle. This is in agreement with the interpretation of
the scalar fields as the logarithm of the Polyakov loop in the
compactified dimension. We prove also that the fundamental constituents
of the theory are
 free fermion excitations corresponding to the winding of
the eigenvalues around their target space, with energy growing
quadratically with the winding number.
As shown by the standard form of the partition functions,
eq. (\ref{partition}),
the states of the theory  are labelled by the irreducible
representations of $U(N)$ (or $SU(N)$), and are
states of $ N $ such free fermions.

The interpretation of QCD2 as a matrix model is natural in a formulation
where the kernel on the cylinder and the partition function on the torus
are expressed as an expansion in exponentials of $ \frac{1}{\tilde{g}^2
{\cal A}} $ rather than in exponentials of $ \tilde{g}^2 {\cal A} $
 as in
the representation expansion. From the mathematical
point of view the bridge, between the two representations is a modular
transformation that generalizes to the cylinder and to the torus the
result already known for the disk \cite{MeOn} .

It would be very interesting to generalize this result to kernels with a
higher number of entries and to arbitrary genus, that is to find for
\eq
{\cal K}_{G,n}(g_1,g_2, \ldots ,g_n;N,t) = \sum_R d_{R}^{2-2G-n}
\chi_{R}(g_1) \ldots \chi_R(g_n)
e^{-\frac{t}{2} C_2 (R) }
\label{highgker}
\en
a representation in terms of exponentials of $ \frac{1}{\tilde{g}^2
 {\cal
A}} $. This would presumably amount to describe such kernels in terms of
one dimensional matrix models whose target space includes both branching
points and loops. There is one obvious difficulty in such program:
by sewing two kernels together one should obtain a new kernel that
depends only on the total area. This is quite natural in
(\ref{highgker}),
as the exponentials are additive in the area, but in principle much more
difficult with expressions containing exponentials of $
\frac{1}{\tilde{g}^2{\cal A}} $ .

\section{Appendix A}

In this appendix we will provide an alternative derivation of the
$SU(N)$ partition function on the cylinder, which follows closely
the method employed by Itzykson and Zuber \cite{IZ} to derive their
well-known formula for the angular integral in hermitian matrix models.

First we note, following \cite{BlTh}, that the partition function on the
cylinder can easily be written in terms of the one on the disk, thanks
to the invariance under area-preserving diffeomorphisms. One just has to
deform the circle into a rectangle, decompose the Wilson loop in the
product of four Wilson lines, and then identify two opposite sides.
Consider then the $ SU(N) $ kernel on the disk, as given by \cite{MeOn},
 before periodicity is imposed:
\eq
{\cal K}_{1}(\phi_i;t) = {\cal N}(t) \prod_{i < j}
\frac{\phi_i - \phi_j}{2 \sin \frac{\phi_i - \phi_j}{2}}
\exp \left[ - \frac{1}{2t} \sum_i \phi_{i}^{2} \right] .
\en
We want to calculate the integral
\eq
{\cal K}_{2}(\phi_{1}^{i} , \phi_{2}^{i} ; t) =
\int_{SU(N)} dh~ {\cal K}_{1}(\Lambda_1 h \Lambda_2 h^{-1} ; t) ,
\en
where $ \Lambda_1 = diag ( e^{i \phi_{1}^{i}} ) ,
\Lambda_2 = diag ( e^{i \phi_{2}^{i}} ) $, and  ${\cal K}_1 $
is a function of the
eigenvalues $e^{i \phi_{U}^{i}}$
of the
unitary matrix $ U = \Lambda_1 h \Lambda_2 h^{-1} $.
To this end,
consider an arbitrary solution of the heat equation on the group,
 subject
to the condition that it should be a class function, and a symmetric
function of the eigenvalues $e^{i \phi_{U}^{i}}$. Denote this function
by
$ f(\phi_{U}^{i} ; t) $. Then
\eqa
f(\phi_U^i ;t) & = &
\int_{SU(N)}~ dg~ {\cal K}_{1} (U g^{-1} , t) f(g,0) \nonumber  \\
 & = & \int_{SU(N)}~ dg~ {\cal K}(U,g;t) f(g;0) .
\ena
Following standard techniques, we integrate first over the eigenvalues
of $g$, obtaining
\eq
f(\phi_{U}^{i};t) = C_N
 \int_{SU(N)} dS \int d\Lambda_g J^2 (\Lambda_g )
{\cal K} (\Lambda_U , S \Lambda_g S^{-1} ; t) f(\Lambda_g ;0) ,
\en
where
\eq
J \left( \Lambda_g \right)
= i \prod_{i < j} 2 \sin \frac{\phi_{(g)}^i - \phi_{(g)}^j}{2} .
\en
Consider now the antisymmetric function of $\Lambda_U$
\eq
\xi (\phi_U^i , t) \equiv \Delta (\Lambda_U ) f( \phi_{U}^i , t) .
\en
We find
\eq
\xi (\phi_U^i ;t) =  \int d\Lambda_g \xi (\phi_g^i , 0)
\hat{\cal{K}} ( \Lambda_g , \Lambda_U ; t) ,
 \en
where the kernel for $\xi$ is
\eq
\hat{\cal{K}} ( \Lambda_g , \Lambda_U ; t) =
C \Delta (\Lambda_g) \Delta (\Lambda_U) \int_{SU(N)}
dS {\cal K} (\Lambda_U , S \Lambda_g S^{-1} ; t) .
\label{Khat}
\en
Since $f(\phi_U^i ,t)$ is a solution of the heat equation , $\xi
(\phi_U^i ,t) $ obeys
\eq
\frac{\partial\xi}{\partial t} = \Delta(\phi_U) \hat{\Delta}_U
f(\phi_U ;t),
\en
where $\hat{\Delta}_U $ is the ``radial'' part of the Laplace-Beltrami
operator on the group manifold, given by \cite{MeOn}
\eq
\hat{\Delta}_U = \frac{1}{2J} \sum_i \frac{\partial^2}{\partial
\phi_i^2}~ J + \frac{1}{2}~ R_N ,
\label{lapbel}
\en
with $ R_N = \frac{1}{12} N(N^2-1)$.
The function $\xi(\phi_U^i , t)$ thus obeys the simple diffusion
equation
\eq
\frac{\partial\xi}{\partial t} = \frac{1}{2} \sum_i
\frac{\partial^2}{\partial \phi_i^2} ~ \xi
+ \frac{1}{2} R_N~ \xi .
\en
For $R_N = 0$, $\xi$ is a solution of a diffusion equation which is
totally antisymmetric in its arguments $\phi_i$, so the corresponding
kernel must be the antisymmetric gaussian
\eqa
\hat{\cal{K}} (\phi_1, \phi_2;t) & = &
\frac{1}{(2\pi t)^{\frac{N-1}{2}}} \frac{1}{N!} \sum_P
(-1)^P \exp \left[ - \frac{1}{2t} \sum_i
\left(\phi_i^{(1)} - \phi_{P(i)}^{(2)} \right)^2 \right] \nonumber
\\
& = & \frac{1}{(2\pi t)^{\frac{N-1}{2}}}  \det
\left[ \exp \left( - \frac{1}{2t} (\phi_{1,i} - \phi_{2,j})^2
\right) \right].
\label{antgaus}
\ena
Equating eq. (\ref{antgaus}) and eq. (\ref{Khat}) we find that our
 integral is given by
\eqa
{\cal K}_{2} (\phi_1^i , \phi_2^i ;t) & = &
\int_{SU(N)} dh {\cal K} (\Lambda_1 , h \Lambda_2 h^{-1};t)
\nonumber  \\
& = & {\cal N}_2(t) \frac{1}{J(\Lambda_1^i) J(\Lambda_2^i)} \det
\left[ \exp \left( -\frac{1}{2t} ( \phi_{1,i} - \phi_{2,j})^2 \right) ,
\right]
\ena
and the only modification due to the $R_N$ term is an extra factor of
$e^{R_N t/2}$, which is also present in \cite{MeOn}.

\newpage
{\bf Acknowledgments}

One of us (M. C.) would like to acknowledge enlightening discussions
with George Thompson.
\vskip 1cm
{\bf Note}

While writing this paper we received two independent papers
{}~\cite{MiPo,Dou} with
new results on QCD2, that partially overlap our work.
In particular the work of Minahan and Polychronakos
{}~\cite{MiPo} was of some help to
us in the interpretation of the grand canonical partition
functions.

\end{document}